\documentclass[12pt]{article}
\def\thebibliography#1{\list{~\arabic{enumi}.}
  {\settowidth\labelwidth{#1.}\leftmargin=2.47em
   \labelsep\leftmargin \advance\labelsep-\labelwidth
   \usecounter{enumi}}\def\makelabel##1{\rlap{##1}\hss}%
   \def\newblock{\hskip 0.11em plus 0.33em minus -0.07em}
   \sloppy \clubpenalty=4000 \widowpenalty=4000 \sfcode`\.=1000\relax}

\newcounter{tempref}
\def \bea{\begin{eqnarray}}

\def \beq{\begin{equation}}

\def \eea{\end{eqnarray}}
\def \eeq{\end{equation}}

\def \ew{SU(2) $\otimes$ U(1)}

\def \ite{{\it et al.}}

\def \pr{\parallel}

\def \U1Y{U(1)$_Y$}

\def \ajp#1#2#3{Am.\ J. Phys.\ {\bf#1}, #2 (#3)}

\def \arnps#1#2#3{Ann.\ Rev.\ Nucl.\ Part.\ Sci.\ {\bf#1}, #2 (#3)}

\def \cn{Collaboration}
\def \cp89{{\it CP Violation,} edited by C. Jarlskog (World Scientific,
Singapore, 1989)}

\def \epjc#1#2#3{Eur.\ Phys.\ J.\ C {\bf#1}, #2 (#3)}
\def \f79{{\it Proceedings of the 1979 International Symposium on Lepton and
Photon Interactions at High Energies,} Fermilab, August 23-29, 1979, ed. by
T. B. W. Kirk and H. D. I. Abarbanel (Fermi National Accelerator Laboratory,
Batavia, IL, 1979}
\def \hb87{{\it Proceeding of the 1987 International Symposium on Lepton and
Photon Interactions at High Energies,} Hamburg, 1987, ed. by W. Bartel
and R. R\"uckl (Nucl.\ Phys.\ B, Proc.\ Suppl., vol. 3) (North-Holland,
Amsterdam, 1988)}

\def \ibj#1#2#3{~{\bf#1}, #2 (#3)}
\def \ichep72{{\it Proceedings of the XVI International Conference on High
Energy Physics}, Chicago and Batavia, Illinois, Sept. 6 -- 13, 1972,
edited by J. D. Jackson, A. Roberts, and R. Donaldson (Fermilab, Batavia,
IL, 1972)}
\def \ijmpa#1#2#3{Int.\ J.\ Mod.\ Phys.\ A {\bf#1}, #2 (#3)}
\def \ite{{\it et al.}}

\def \lkl87{{\it Selected Topics in Electroweak Interactions} (Proceedings of
the Second Lake Louise Institute on New Frontiers in Particle Physics, 15 --
21 February, 1987), edited by J. M. Cameron \ite~(World Scientific, Singapore,
1987)}
\def \kaon{{\bf Kaon Physics}, edited by J. L. Rosner and B. Winstein
(University of Chicago Press, 2001)}

\def \nc#1#2#3{Nuovo Cim.\ {\bf#1}, #2 (#3)}
\def \nima#1#2#3{Nucl.\ Instr.\ Meth.\ A {\bf#1}, #2 (#3)}
\def \np#1#2#3{Nucl.\ Phys.\ {\bf#1}, #2 (#3)}
\def \npps#1#2#3{Nucl.\ Phys.\ Proc.\ Suppl.\ {\bf#1}, #2 (#3)}

\def \pisma#1#2#3#4{Pis'ma Zh.\ Eksp.\ Teor.\ Fiz.\ {\bf#1}, #2 (#3) [JETP
Lett.\ {\bf#1}, #4 (#3)]}
\def \pl#1#2#3{Phys.\ Lett.\ {\bf#1}, #2 (#3)}

\def \plb#1#2#3{Phys.\ Lett.\ B {\bf#1}, #2 (#3)}
\def \ppmsj#1#2#3{Proc.\ Phys.\ Math.\ Soc.\ Japan {\bf#1}, #2 (#3)}
\def \pr#1#2#3{Phys.\ Rev.\ {\bf#1}, #2 (#3)}

\def \prd#1#2#3{Phys.\ Rev.\ D {\bf#1}, #2 (#3)}
\def \prl#1#2#3{Phys.\ Rev.\ Lett.\ {\bf#1}, #2 (#3)}
\def \prp#1#2#3{Phys.\ Rep.\ {\bf#1}, #2 (#3)}
\def \ptp#1#2#3{Prog.\ Theor.\ Phys.\ {\bf#1}, #2 (#3)}
\def \rmp#1#2#3{Rev.\ Mod.\ Phys.\ {\bf#1}, #2 (#3)}

\def \sci#1#2#3{Science {\bf#1}, #2 (#3)}
\def \si90{25th International Conference on High Energy Physics, Singapore,
Aug. 2-8, 1990}
\def \slc87{{\it Proceedings of the Salt Lake City Meeting} (Division of
Particles and Fields, American Physical Society, Salt Lake City, Utah, 1987),
ed. by C. DeTar and J. S. Ball (World Scientific, Singapore, 1987)}
\def \slac89{{\it Proceedings of the XIVth International Symposium on
Lepton and Photon Interactions,} Stanford, California, 1989, edited by M.
Riordan (World Scientific, Singapore, 1990)}
\def \smass82{{\it Proceedings of the 1982 DPF Summer Study on Elementary
Particle Physics and Future Facilities}, Snowmass, Colorado, edited by R.
Donaldson, R. Gustafson, and F. Paige (World Scientific, Singapore, 1982)}
\def \smass90{{\it Research Directions for the Decade} (Proceedings of the
1990 Summer Study on High Energy Physics, June 25--July 13, Snowmass,
Colorado),
edited by E. L. Berger (World Scientific, Singapore, 1992)}

\def \TASI{{\bf TASI-2000:  Flavor Physics for the Millennium}, edited by J. L.
Rosner (World Scientific, Singapore, 2001)}
\def \yaf#1#2#3#4{Yad.\ Fiz.\ {\bf#1}, #2 (#3) [Sov.\ J.\ Nucl.\ Phys.\
{\bf #1}, #4 (#3)]}
\def \zhetf#1#2#3#4#5#6{Zh.\ Eksp.\ Teor.\ Fiz.\ {\bf #1}, #2 (#3) [Sov.\
Phys.\ - JETP {\bf #4}, #5 (#6)]}

\begin{document}
\rightline{EFI 02-89-Rev}
\rightline{hep-ph/0206176}
\rightline{January 2003}
\medskip

\centerline{\bf \large Resource Letter SM-1:  The Standard Model and Beyond}
\medskip
\centerline{Jonathan L. Rosner}
\smallskip
\centerline{\it Enrico Fermi Institute and Department of Physics}
\centerline{\it University of Chicago,
5640 South Ellis Avenue, Chicago IL 60637, USA} 

\begin{quote}
This Resource Letter provides a guide to literature on the Standard Model
of elementary particles and possible extensions.  In the successful theory of
quarks and leptons and their interactions, important questions remain, such as
the mechanism of electroweak symmetry breaking, the origin of quark and lepton
masses, the source of the baryon asymmetry of the Universe, and the makeup of
its matter and energy density.  References are cited for quarks and leptons,
gauge theories, color and chromodynamics, weak interactions, electroweak
unification, CP violation, dynamics of heavy quarks, Higgs bosons,
precision electroweak measurements, supersymmetry, dynamical electroweak
symmetry breaking, composite quarks and leptons, grand unification and extended
gauge groups, string theories, large extra dimensions, neutrino masses, cosmic
microwave background radiation, dark matter, dark energy, accelerator
facilities, and non-accelerator experiments.
\end{quote}
\bigskip

\leftline{\bf I.  INTRODUCTION}
\bigskip

The ``Standard Model'' of elementary particle physics encompasses the progress
of the past half-century in understanding the weak, electromagnetic, and strong
interactions.  During this period tremendous strides were made in bringing
quantum field theory to bear upon a wide variety of phenomena.

The arsenal of techniques for understanding the strong interactions in the
1960s included principles based on analyticity, unitarity, and symmetry.  The
successes of the emerging quark model often seemed mysterious.  The ensuing
decade yielded a theory of strong interactions, quantum chromodynamics (QCD),
permitting calculations of a wide range of properties of the {\it hadrons}, or
strongly interacting particles, and has been validated by the discovery of its
force-carrier, the {\it gluon}.

In the 1960s the weak interactions were represented by a phenomenological
four-fermion theory of no use for higher-order calculations. Attempts to
describe weak interactions with heavy boson exchange bore fruit when these
interactions were unified with electromagnetism and a suitable mechanism for
generation of heavy boson mass was found.  This {\it electroweak theory} has
been spectacularly successful, leading to the prediction and observation of the
$W$ and $Z$ bosons and to precision tests confirming the theory's validity in
higher-order calculations.

This Resource Letter begins with sections devoted to the resources available
for study of the Standard Model of particle physics and its extensions:
periodicals (II), conference proceedings (III), texts and reviews (IV),
historical references (V), popular literature (VI), Internet resources (VII),
and a guide to Nobel prizes related to the subject (VIII).

A description of Standard Model research literature follows.  In Section IX,
based in part on \cite{Rosner:2001zy}, the
ingredients of the standard model --- the quarks and leptons and their
interactions -- are introduced, and QCD is discussed briefly.  The unified
theory of weak and electromagnetic interactions is described, its role in
explaining CP violation is explained, and its missing piece -- the Higgs
boson -- is mentioned.

Important questions remain that are not addressed in the Standard Model.
These include the unification of the electroweak and strong interactions
(possibly including gravity), the origin of quark and lepton masses, the source
of the baryon asymmetry of the Universe, and the nature of its unseen matter
and energy density.  Some proposed Standard Model extensions devoted to these
problems are noted in Section X.  Concrete evidence for physics beyond the
Standard Model, including neutrino masses, cosmic microwave background
radiation, dark matter, and ``dark energy,'' is described in Section XI.  A
variety of experimental methods are appropriate for probing these phenomena
(Section XII).  A brief summary (Section XIII) concludes.

\leftline{\bf II.  PERIODICALS}
\bigskip

The literature on the Standard Model of particle physics and its extensions
is extensive and international, but a good sense of the field can be gained
by perusing about a dozen main journals.  Subsequent sections are devoted to
other means of gaining information about this rapidly changing subject.
\bigskip

\noindent
Instrumentation journals with some articles on elementary particle physics:

{\it IEEE Transactions on Nuclear Science}

{\it Nuclear Instruments and Methods A}

{\it Review of Scientific Instruments}

\noindent
Journals devoted primarily or largely to elementary particle physics:

{\it European Journal of Physics C}

{\it Fizika Elementarnykh Chastits i Atomnogo Yadra} (Soviet Journal of
Particles and \\ \null \qquad \qquad Nuclei)

{\it International Journal of Modern Physics A}

{\it Journal of High Energy Physics} (``JHEP''; electronic)

{\it Journal of Physics G}

{\it Modern Physics Letters A}

{\it Nuclear Physics B}

{\it Nuovo Cimento A}

{\it Physical Review D}

{\it Physics Letters B}

{\it Progress of Theoretical Physics (Kyoto)}

{\it Yadernaya Fizika} (Soviet Journal of Nuclear Physics --1992; Physics of
Atomic \\ \null \qquad \qquad Nuclei 1993--).

{\it Zeitschrift f\"ur Physik C, now absorbed into European Journal of Physics
C}

{\it Zhurnal Eksperimental'nyi i Teoreticheskii Fizika (Soviet Physics -
JETP)}

\noindent
Laboratory newsletters:

{\it CERN Courier} (European Center for Nuclear Research); web address:\\
\null \qquad \qquad {\tt http://www.cerncourier.com/}

{\it FermiNews} (Fermilab, USA); web address:\\
\null \qquad \qquad {\tt http://www.fnal.gov/pub/ferminews/}

{\it SLAC Beam Line} (Stanford Linear Accelerator Center); web address:\\
\null \qquad \qquad {\tt http://www.slac.stanford.edu/pubs/beamline/}

\noindent
Rapid publication journals with section devoted to particle physics:

{\it Chinese Physics Letters}

{\it Europhysics Letters}

{\it Physical Review Letters}

{\it Pis'ma v Zhurnal Eksperimental'nyi i Teoreticheskii Fizika
(JETP Letters)}

\noindent
Review journals:

{\it Annals of Physics (N.Y.)}

{\it Annual Review of Nuclear and Particle Science}

{\it Physics Reports}

{\it Reports on Progress in Physics}

{\it Reviews of Modern Physics}

\noindent
Other journals with frequent articles on particle physics or related subjects:

{\it Acta Physica Polonica}

{\it American Journal of Physics}

{\it Astroparticle Physics}

{\it Astrophysical Journal}

{\it Nature}

{\it New Scientist}

{\it Physics Today (AIP)}

{\it Physics World (IOP)}

{\it Progress of Theoretical Physics (Japan)}

{\it Science}

{\it Science News}

{\it Scientific American}

\bigskip

\leftline{\bf III.  CONFERENCE PROCEEDINGS}
\bigskip

The latest biennial ``Rochester'' Conference in High Energy Physics was held in
Amsterdam in July 2002; the previous one was in Osaka in 2000 \cite{ICHEP2000}.
In odd-numbered years there occur both the International Symposium on
Lepton and Photon Interactions at High Energies, of which the most recent
was in Rome \cite{LG01}, and the International Europhysics Conference on
High Energy Physics, most recently held in Budapest \cite{EHEP}.  The
locations of each of these conferences since 1990 are summarized in
Table \ref{tab:confs}.  A search of the SPIRES listing at the SLAC Library
(see Sec.\ VII) is the easiest way to find the corresponding Proceedings.
\bigskip

\begin{table}[h]
\caption{Locales of major high energy physics conferences since 1990.
(1) International Conference on High Energy Physics (``Rochester''
Conference); (2) International Symposium on Lepton and Photon Interactions at
High Energies; (3) International Europhysics Conference on High Energy Physics. 
\label{tab:confs}}
\begin{center}
\begin{tabular}{|c c|c c|c c|} \hline \hline
\multicolumn{2}{|c}{(1)} & \multicolumn{2}{|c}{(2)} & \multicolumn{2}{|c|}{(3)}
\\
Year & Location & Year & Location & Year & Location \\ \hline
1990 & Singapore & 1991 & Geneva & 1991 & Geneva \\
1992 & Dallas, TX & 1993 & Ithaca, NY & 1993 & Marseille \\
1994 & Glasgow & 1995 & Beijing & 1995 & Brussels \\
1996 & Warsaw & 1997 & Hamburg & 1997 & Jerusalem \\
1998 & Vancouver & 1999 & Stanford & 1999 & Tampere, Finl.\ \\
2000 & Osaka & 2001 & Rome & 2001 & Budapest \\
2002 & Amsterdam & 2003 & Fermilab & 2003 & Aachen \\ \hline \hline
\end{tabular}
\end{center}
\end{table}

\leftline{\bf IV.  TEXTBOOKS, EXPOSITIONS, AND REVIEW ARTICLES}
\bigskip

This section indicates textbooks and articles at the intermediate or
advanced level.  For popularizations at the non-specialist's level, see
Section VI.
\bigskip

\leftline{\bf A.  Textbooks}
\bigskip

\noindent
{\it 1.  Quantum field theory:}

\noindent
{\it 2.  Standard model (electroweak and strong interactions):}

\noindent
{\it 3.  CP violation:}

\noindent
{\it 4.  Elementary particle phenomenology:}

\noindent
{\it 5.  Symmetries:}

\noindent
{\it 6.  Higgs boson(s):}

\noindent
{\it 7.  Neutrinos:}

\noindent
{\it 8.  Supersymmetry:}

\noindent
{\it 9.  Beyond the Standard Model:}

\noindent
{\it 10.  String theory:}

\bigskip

\leftline{\bf B.  Expositions (summer school lectures, collections of
articles)}
\bigskip

A conference on kaon physics was held at the University of Chicago in 1999
as part of a series.  A volume of articles based on the conference gives
an overview of the field \cite{Kaon99}.

Regular summer schools in particle physics are organized in several locales,
including Boulder (Colorado), Carg\`ese (Corsica), CERN, Erice (Sicily),
and SLAC (Stanford).  The topics typically vary from year to year but there
are frequently lectures on various aspects of the Standard Model (see, e.g.,
the lectures on CP violation by Nir \cite{Nir:1999mg} and the overview by
Rosner \cite{Rosner:2001zy}).

The Theoretical Advanced Study Institute (TASI) at the University of
Colorado was devoted in June of 2000 to flavor physics, a major aspect
of the Standard Model, and the proceedings also contain various aspects of
proposed physics beyond the Standard Model \cite{TASI2000}.  For specific
reviews given at summer schools, see the subsection on Reviews, below.

\leftline{\bf C.  Review Articles}
\bigskip

A number of review articles will be referred to in the narrative of the
Standard Model and its extensions (Secs.\ X--XIII).  These include
the following:

\medskip
\noindent
{\it 1.  Gauge theories:}

\noindent
{\it 2.  Standard Model:} In addition to Rosner (2001) \cite{Rosner:2001zy},
see:

\noindent
{\it 3. Hadron spectra and quarks:}

\noindent
{\it 4.  Group theory:}

\noindent
{\it 5.  Neutrino physics}:
\bigskip

\noindent
Massive neutrinos and neutrino oscillations:

\noindent
Precision electroweak measurements using neutrinos:

\noindent
{\it 6.  Supersymmetry:}

\noindent
{\it 7.  Extended gauge theories:}

\noindent
{\it 8.  Atomic parity violation:}

\noindent
{\it 9.  Particle properties and general lore:}

\medskip
\noindent
A wide variety of mini-reviews of various aspects of the Standard Model may be
found in the Review of Particle Physics published by the Particle Data Group:

\leftline{\bf D.  Other Resource Letters}

\leftline{\bf V.  HISTORICAL REFERENCES}
\bigskip

A symposium on the history of Symmetries in Physics from 1600 to 1980
\cite{Doncel:1983} contains many informative articles.  For a series of
conferences on the history of particle physics, culminating in the rise of the
Standard Model, see \cite{Brown:1984rs,Brown:1989im,Brown:1997}.  The history
of quantum electrodynamics is detailed in \cite{Schweber:1994qa}, while
Pais \cite{Pais:1986} has chronicled the development of particle physics
with particular emphasis on its earlier aspects.  A review of some later
developments is given in \cite{Fitch:1994cq}.  Personal memoirs include those
of a theorist with close ties to experiment (Sam B. Treiman
\cite{Treiman:1996ap}) and a Nobel-prize-winning experimentalist
(Jack Steinberger \cite{Steinberger:1997}).  A collection of articles on
supersymmetry with a historical flavor is based on a recent symposium
\cite{Kane:2000ew}.  Two excellent accounts of experimental high energy physics
by P. Galison are \cite{Galison:1987gr} and \cite{Galison:1997hg}.

\leftline{\bf VI.  POPULAR LITERATURE}
\bigskip

\leftline{\bf A.  Books}
\bigskip

For descriptions of particle theory in a cosmological context see
\cite{Weinberg:1977,Pagels:1982}.  A well-written account of the experiments
that led to the idea of quarks being taken seriously is given in
\cite{Riordan:1987gw}.  The goals of particle theory are described in
\cite{Wilczek:1988ke,Kane:1995,Hooft:1997}, while \cite{Greene:1999kj%
,Weinberg:1992nd} give the
case for a fully unified theory.  The ongoing search for the Higgs particle
and many other efforts in particle physics are treated by
\cite{Lederman:1993nx}.  Gordon Fraser, the former editor of the {\it CERN
Courier}, has written or edited several fine books on particle physics aimed at
general audiences \cite{Fraser:1995,Fraser:1997,Fraser:1998,Fraser:2000}.
One recent popular book on quantum mechanics has been written by Sam Treiman
\cite{Treiman:1999wh}.  Many fine popularizations have been written by
Richard P. Feynman, including his book on quantum electrodynamics
\cite{Feynman:1985} and his Dirac Memorial Lecture, jointly in a volume with
that by Steven Weinberg \cite{Feynman:1987}.

\leftline{\bf B.  Articles}
\bigskip

Instructive popular articles (in more or less chronological order) include ones
by Lederman on the discovery of the Upsilon particle (the first evidence for
the $b$ quark) \cite{Lederman:1978gi},
't Hooft on gauge theories \cite{'tHooft:1980us},
Wilczek \cite{Wilczek:1980vy} and Quinn and Witherell \cite{Quinn:1998ru}
on matter-antimatter asymmetry,
Georgi on quark-lepton and strong-electroweak unification \cite{Georgi:1981yp},
Weinberg \cite{Weinberg:1981kk}, Losecco {\it et al.} \cite{Losecco:1985xj},
and Langacker \cite{Langacker:1992qb} on proton decay,
Quigg on elementary particles and forces \cite{Quigg:1985ai},
Haber and Kane on supersymmetry \cite{Haber:1986ht},
Veltman on the Higgs boson \cite{Veltman:1986br},
Krauss on dark matter in the Universe \cite{Krauss:1986mj},
Green \cite{Green:1986xd} and Duff \cite{Duff:1998ec} on string theory,
Rees on the Stanford Linear Collider \cite{Rees:1989zb},
Bahcall on the solar neutrino problem \cite{Bahcall:1990qf},
Myers and Picasso on the LEP Collider at CERN \cite{Myers:1990gu},
Lederman on the Fermilab Tevatron \cite{Lederman:1991tm},
Feldman and Steinberger on measurements at LEP and SLC suggesting the
existence of three families of quarks and leptons \cite{Feldman:1991kw},
Liss and Tipton on the discovery of the top quark \cite{Liss:1997vk},
Hogan {\it et al.} on supernova surveys and the
accelerating Universe \cite{Hogan:1999mx},
Kearns {\it et al.} on detecting massive neutrinos \cite{Kearns:1999kr},
Weinberg on the goal of a truly unified theory \cite{Weinberg:1999az}
(see below for an Internet link on this article),
Llewellyn Smith on the Large Hadron Collider \cite{LlewellynSmith:2000nb},
Caldwell and Kamionkowski \cite{Caldwell:2001kq} and Gibbs \cite{Gibbs:2002js}
on the cosmic microwave background radiation,
Ostriker and Steinhardt on ``dark energy'' \cite{Ostriker:2001ik}, and
Arkani-Hamed {\it et al.} \cite{Arkani-Hamed:2000,Arkani-Hamed:2002rz} on
large extra dimensions.  {\it The Economist} carries frequent and
well-informed articles on progress in high energy physics (see, e.g.,
\cite{Economist:2002}).  Shorter news articles appear regularly in {\it
Nature}, {\it Science}, and {\it Scientific American}.

\leftline{\bf VII.  INTERNET RESOURCES}
\bigskip

\leftline{\bf A.  Preprints}
\bigskip

A comprehensive repository of preprints on experimental and theoretical
particle physics may be found at {\tt http://arXiv.org/}, including
experimental papers at {\tt http://arXiv.org/archive/hep-ex}, phenomenological
papers (theory papers dealing with experiment) at
{\tt http://arXiv.org/archive/hep-ph},
and more abstract theoretical papers at {\tt http://arXiv.org/archive/hep-th}.
The SPIRES system at Stanford Linear Accelerator Center:
{\tt http://www.slac.stanford.edu/spires/} lists a number of different
categories, including books, conferences, experiments, preprints (SPIRES HEP),
and even names and e-mail addresses of particle physicists.
\bigskip

\leftline{\bf B.  Laboratories and accelerators}
\bigskip

National and international high energy physics maintain extensive web pages
with vast links to useful information.  For a comprehensive listing, see \\
{\tt http://www.nevis.columbia.edu/\~{}quarknet/%
high\_energy\_physics\_links.htm}. \\
Some examples are given in Tables \ref{tab:lablinks} and \ref{tab:nalinks}.

\begin{table}[h]
\caption{Major accelerator-based HEP laboratories and their public web pages.
\label{tab:lablinks}}
\begin{center}
\begin{tabular}{l l c} \hline \hline
Laboratory & Location & Web address \\ \hline
Brookhaven & Upton, New York, USA & {\tt http://www.bnl.gov/world/} \\
Budker Inst.\ & Novosibirsk, Russia & 
{\tt http://www.inp.nsk.su/index.en.shtml} \\
CERN & Geneva, Switz.\ & {\tt http://public.web.cern.ch/Public/} \\
Cornell & Ithaca, New York, USA & {\tt http://www.lns.cornell.edu} \\
DESY & Hamburg, Germany & {\tt http://www.desy.de/html/home/} \\
Fermilab & Batavia, IL, USA & {\tt http://www.fnal.gov/} \\
Frascati & Frascati, Italy & {\tt http://www.lnf.infn.it/} \\
IHEP & Beijing, China & {\tt http://www.ihep.ac.cn/} \\
IHEP & Protvino, Russia & {\tt http://www.ihep.su/} \\
KEK & Tsukuba, Japan & {\tt http://www.kek.jp/intra.html} \\
SLAC & Stanford, Calif., USA & {\tt http://www.slac.stanford.edu} \\
TJNAF & Newport News, VA, USA & {\tt http://www.jlab.org/} \\
\hline \hline
\end{tabular}
\end{center}
\end{table}

\begin{table}[h]
\caption{Major non-accelerator laboratories and their public web pages.
\label{tab:nalinks}}
\begin{center}
\begin{tabular}{l l c} \hline \hline
Laboratory & Location & Web address \\ \hline
Gran Sasso & Central Italy & {\tt http://www.lngs.infn.it/} \\
Kamioka & Western Japan & {\tt http://www-sk.icrr.u-tokyo.ac.jp/} \\
Soudan & Northern Minn.\ & {\tt http://www.hep.umn.edu/soudan/} \\
Sudbury $\nu$ Obs.\ & Ontario & {\tt http://www.sno.phy.queensu.ca/}
 \\ \hline \hline
\end{tabular}
\end{center}
\end{table}

The site {\tt http://physics.web.cern.ch/Physics/HEPWebSites.html} contains a
number
of links to further web pages, including the CERN Large Hadron Collider at \\
{\tt http://lhc-new-homepage.web.cern.ch/lhc-new-homepage/}, the ``Particle
Adventure'' site {\tt http://particleadventure.org/particleadventure} of
the Particle Data Group at Lawrence Berkeley National Laboratory, and
{\it Quarknet}, a network for high school science
teachers to involve them and their students in cutting-edge research in
particle physics at {\tt http://quarknet.fnal.gov/}.  The IHEP laboratory
in Russia hosts a chronology of particle physics discoveries:\\
{\tt http://ontil.ihep.su/\~{}ppds/discovery.html}.
\bigskip

\leftline{\bf C.  Popular article with extensive links}
\bigskip

The {\bf Scientific American} article on the future of particle
physics by Steven Weinberg \cite{Weinberg:1999az} appears on the web with a
variety of links to other literature:

\noindent
{\tt http://www.sciam.com/issue.cfm?issueDate=Dec-99}.
\bigskip

\leftline{\bf VIII.  NOBEL PRIZES RELATED TO THE STANDARD MODEL}
\bigskip

Some contributions in the past 45 years related to the formulation of
the Standard Model that have been recognized by Nobel Prizes in Physics are
summarized in Table \ref{tab:nob}.  More information may be found on
the web sites {\tt http://www.slac.stanford.edu/ \\ library/nobel.html}
and {\tt http://www.nobel.se/physics/laureates}.  Many
additional prizes were awarded for instrumentation or discoveries crucial to
our present understanding of the Standard Model.
\bigskip

\begin{table}[h]
\caption{Nobel prizes in physics since 1957 related to the Standard Model.
\label{tab:nob}}
\begin{center}
\begin{tabular}{l l l} \hline \hline
Year & Recipient(s) & Subject \\ \hline
1957 & T. D. Lee and C. N. Yang & Parity violation \\
1960 & D. A. Glaser & Bubble chamber \\
1965 & R. P. Feynman, J. S. Schwinger, & \\
     & \quad and S. I. Tomonaga & Quantum electrodynamics \\
1968 & L. W. Alvarez & Discovery of resonances \\
1969 & M. Gell-Mann & Particle classification \\
1976 & B. Richter and S. C. C. Ting & $J/\psi$ discovery \\
1979 & S. L. Glashow, A. Salam, & \\
     & \quad and S. Weinberg & Electroweak unification \\
1980 & J. W. Cronin and V. L. Fitch & CP violation \\
1982 & K. G. Wilson & Critical phenomena \\
1984 & C. Rubbia and & $W$ and $Z$ discovery via\\
     & \quad S. Van Der Meer & \quad S$\bar p p$S collider \\
1988 & L. M. Lederman, M. Schwartz, & Discovery that \\
     & \quad and J. Steinberger & \quad $\nu_\mu \ne \nu_e$ \\
1990 & J. I. Friedman, H. W. Kendall, & Deep inelastic electron \\
     & \quad and R. E. Taylor & \quad scattering \\
1992 & G. Charpak & Particle detectors \\
1995 & M. L. Perl & $\tau$ lepton \\
     & F. Reines & Neutrino detection \\
1999 & G. 't Hooft and & \\
     & \quad M. J. G. Veltman & Electroweak interactions \\
2002 & R. Davis and M. Koshiba & Cosmic neutrinos \\
     & R. Giacconi & Cosmic X-rays \\
\hline \hline
\end{tabular}
\end{center}
\end{table}
\bigskip

\leftline{\bf IX.  SNAPSHOT OF THE STANDARD MODEL}
\bigskip

\leftline{\bf A.  Quarks and leptons}
\bigskip


The major ingredients of the Standard Model have been in place for some time,
and can be gleaned from the popular article by Quigg \cite{Quigg:1985ai}.
The known building blocks of strongly interacting particles, the {\it quarks}
\cite{Gell-Mann:1964nj,Zweig:1981pd,Zweig:1964jf}, and the fundamental fermions
lacking strong interactions, the {\it leptons}, are
summarized in Table
\ref{tab:ql}.  The quark masses quoted there \cite{Hagiwara:2002pw} are those
for quarks probed at distances short compared with the characteristic
size of strongly interacting particles.  When regarded as constituents of
strongly interacting particles, however, the $u$ and $d$ quarks act as
quasi-particles with masses of about 0.3 GeV.  The corresponding
``constituent-quark'' masses of $s$, $c$, and $b$ are about 0.5, 1.5, and 4.9
GeV, respectively \cite{Gasiorowicz:1981jz}.  (For reviews of the spectroscopy
of hadrons containing the heavy quarks $c$ and $b$, see
\cite{Appelquist:1978aq,Novikov:1978dq,Kwong:1987mj,Buchmuller:1988nq}.)  The
pattern of charge-changing weak transitions
between quarks with charges $Q = 2/3$ and those with charges $Q = -1/3$
is described by the $3 \times 3$ {\it Cabibbo-Kobayashi-Maskawa}
\cite{Cabibbo:1963yz,Kobayashi:1973fv}, or {\it CKM} matrix; for a review of
its properties, see \cite{Gilman:2002}.

\begin{table}
\caption{The known quarks and leptons.  Masses in GeV except where
indicated otherwise.  Here and elsewhere $c=1$.
\label{tab:ql}}
\begin{center}
\begin{tabular}{|c c|c c|c c|c c|} \hline
\multicolumn{4}{|c}{Quarks} & \multicolumn{4}{|c|}{Leptons} \\ \hline
\multicolumn{2}{|c}{Charge $2/3$} & \multicolumn{2}{|c}{Charge $-1/3$} &
\multicolumn{2}{|c}{Charge $-1$} & \multicolumn{2}{|c|}{Charge 0} \\ \hline
    & Mass &      & Mass &      & Mass &     & Mass \\ \hline
$u$ & 0.0015--0.0045 & $d$ & 0.005--0.0085 & $e$ & 0.000511 & $\nu_e$ & 
 $< 3$ eV \\
$c$ & 1.0--1.4 & $s$ & 0.085--0.155 & $\mu$ & 0.106 & $\nu_\mu$ & 
 $<190$ keV \\
$t$ & $174.3 \pm 5.1$ & $b$ & 4.0--4.5 & $\tau$ & 1.777 & $\nu_\tau$ &
 $<18.2$ MeV \\ \hline
\end{tabular}
\end{center}
\end{table}

The quarks and leptons in Table \ref{tab:ql} fall into three ``families.''
For evidence that all the existing families (at least those containing light
neutrinos) may have been discovered, see \cite{Feldman:1991kw}.

\leftline{\bf B.  Gauge theories}
\bigskip

A theory of particles and their interactions permitting arbitrary changes of
phase in the particle's quantum mechanical state is an {\it Abelian local gauge
theory} such as electromagnetism.  The term ``Abelian'' indicates that gauge
(phase) transformations at a given space-time point commute with one another,
while ``local'' stands for the freedom to make separate gauge transformations
at each space-time point.  The name ``gauge'' originated with Hermann Weyl
\cite{Weyl:1929fm}.

Gauge transformations may be generalized to those that do {\it not} commute
with one another at a given space-time point.  The first such {\it
non-Abelian} gauge theory was proposed by C. N. Yang and R. L. Mills
\cite{Yang:1954ek}, who used it to describe the strong interactions through
self-interacting mesons of spin 1 carrying isosopic spin.

The review by Abers and Lee \cite{Abers:1973qs} helped a generation of
physicists to apply gauge theories to the electroweak and strong
interactions.  An excellent introduction to the subject at the intermediate
graduate level is given by Quigg \cite{Quigg:1997}.  An article addressed to
the lay reader has been written by 't Hooft \cite{'tHooft:1980us}.  A recent
text \cite{O'Raifeartaigh:1997ia} provides a further introduction to the
subject.

\leftline{\bf C.  Color and quantum chromodynamics}
\bigskip

The quarks are distinguished from the leptons by possessing a three-fold
charge known as ``color'' that enables them to interact strongly with one
another \cite{Greenberg:1964pe,Nambu:1966,Fritzsch:1973pi}. We also speak
of quark and lepton ``flavor'' when distinguishing the particles in Table
\ref{tab:ql} from one another.  The evidence for color comes from several
quarters.

{\em 1.  Quark statistics.}  The $\Delta^{++}$, a low-lying excited state of
the nucleon, can be represented in the quark model as $uuu$, so it is totally
symmetric in flavor.  It has spin $J = 3/2$, a totally symmetric combination of
the three $J=1/2$ quark spins.  As a ground state, its spatial wave function
should be symmetric as well.  While a state composed of fermions should be
totally {\it antisymmetric} under the interchange of any two fermions, the
state described so far is totally {\it symmetric} under the product of flavor,
spin, and space interchanges.  Color introduces an additional degree of freedom
under which the interchange of two quarks can produce a minus sign.

{\em 2.  Electron-positron annihilation to hadrons.}  The charges of all
quarks that can be produced in pairs at a given center-of-mass energy
is measured by the ratio $R \equiv \sigma(e^+ e^- \to {\rm hadrons})/
\sigma(e^+ e^- \to \mu^+ \mu^-) = \sum_i Q_i^2$, where $Q_i$ is the charge of
quark $i$ in units of $|e|$.  Measurements \cite{Hagiwara:2002pw} indicate
values of $R$ in various energy ranges consistent with $N_c = 3$ (with a small
positive correction associated with the strong interactions of the quarks).

{\em 3.  Neutral pion decay.}  The $\pi^0$ decay rate is governed by a quark
loop diagram in which two photons are radiated by the quarks in $\pi^0 =
(u \bar u - d \bar d)/\sqrt{2}$.  The predicted rate is $\Gamma(\pi^0 \to
\gamma \gamma) = 7.6 S^2$ eV, where $S = N_c(Q_u^2 - Q_d^2) = N_c/3$.  The
experimental rate is $7.8 \pm 0.6$ eV, in accord with experiment if $S=1$ and
$N_c = 3$.

{\em 4.  Triality.}  Quark composites appear only in multiples of three.
Baryons are composed of $qqq$, while mesons are $q \bar q$ (with total
quark number zero).  This is compatible with our current understanding
of QCD, in which only color-singlet states can appear in the spectrum.

A crucial feature of the QCD theory of strong interactions is its ``asymptotic
freedom,'' a weakening interaction strength at short distances permitting the
interpretation of deep inelastic scattering experiments \cite{Riordan:1987gw,%
Bloom:1969kc,Breidenbach:1969kd} in terms of quarks.  This property was found
to be characteristic of non-Abelian gauge theories such as color SU(3) by Gross
and Wilczek \cite{Gross:1973id,Gross:1973ju,Gross:1974cs} and by Politzer
\cite{Politzer:1973fx,Politzer:1974fr}.  The result was obtained earlier for
the gauge group
SU(2) by Khriplovich \cite{Khriplovich:1969aa} (see also \cite{'t Hooft:1972}),
but its significance for a strong-interaction theory was not realized then. 
 
Direct evidence for the quanta of QCD, the gluons, was first presented in
1979 on the basis of extra ``jets'' of particles produced in electron-positron
annihilations to hadrons.  Normally one sees two clusters of energy associated
with the fragmentation of each quark in $e^+ e^- \to q \bar q$ into hadrons.
However, in some fraction of events an extra jet was seen, corresponding to
the radiation of a gluon by one of the quarks.  For a popular history of
this discovery, containing further references, see \cite{Riordan:1987gw}.

The transformations that take one color of quark into another are those of
the group SU(3).  This group is called SU(3)$_{\rm color}$
to distinguish it from the SU(3)$_{\rm flavor}$ associated with the quarks $u$,
$d$, and $s$.

\leftline{\bf D.  Weak interactions}
\bigskip

The electromagnetic interaction is described in terms of photon exchange.
The quantum electrodynamics of photons and electrons initially encountered
divergent quantities tamed in the 1940s through {\it renormalization}, leading
to successful estimates of the anomalous magnetic moment of the electron and
the Lamb shift in hydrogen \cite{Schweber:1994qa}.
By contrast, the weak interactions as formulated up to the mid-1960s involved
the pointlike interactions of two currents.  This interaction is very singular
and cannot be renormalized.  The weak currents in this theory were purely
charge-changing.  As a result of work by Gershtein and Zel'dovich (who
suggested that the weak vector current is of universal strength)
\cite{GZ1956}, Lee and Yang \cite{Lee:1956qn,Lee:1957qr,Lee:1957qq},
Feynman and Gell-Mann \cite{Feynman:1958ty}, and Sudarshan and Marshak
\cite{Sudarshan:1958vf}, the weak currents were identified as having
(vector)--(axial) or ``$V-A$'' form.

\leftline{\bf E.  Electroweak unification}
\bigskip

Yukawa \cite{Yukawa:1935xg} and Klein \cite{Klein:1986jm} proposed early
boson-exchange models for the charge-changing weak interactions. Klein's model
had self-interacting bosons, thus anticipating the theory of Yang and Mills
\cite{Yang:1954ek}.  Schwinger and others studied such models in the 1950s, but
Glashow \cite{Glashow:1961tr} realized that a new {\it neutral} heavy boson
$Z$, in addition to the massless photon and massive charged bosons, was needed
to successfully unify the weak and electromagnetic interactions. The use of the
Higgs \cite{Higgs:1964ia,Higgs:1964pj,Englert:1964et,Guralnik:1964eu}
mechanism to break the electroweak symmetry by Weinberg \cite{Weinberg:1967tq}
and Salam \cite{Salam:1968rm} converted this phenomenological theory into one
suitable for higher-order calculations.

The charge-changing weak currents could be viewed as
members of an SU(2) algebra \cite{Gell-Mann:1960np,Cabibbo:%
1963yz}. However, the neutral member of this multiplet could not be identified
with electric charge.  Charged $W^\pm$ bosons couple only to left-handed
fermions, while the photon couples to both left and right-handed fermions.
Moreover, a theory with only photons and charged weak bosons leads to
unacceptable divergences in higher-order processes \cite{Quigg:1997}.  The
neutral heavy $Z$ boson can be arranged to cancel these divergences.  It leads
to {\it neutral current interactions}, in which (for example) an incident
neutrino scatters inelastically on a hadronic target without changing its
charge.  The discovery of neutral-current interactions of neutrinos
\cite{Hasert:1973cr,Hasert:1973ff,Hasert:1974ju,Benvenuti:1974rt} and
other manifestations of the $Z$ strikingly confirmed the new theory.

A key stumbling block to the construction of an electroweak theory applying
to the quarks known at the time ($u$, $d$, and $s$) was the presence of
{\it flavor-changing neutral currents}.  The hypothesis of a fourth ``charmed''
quark $c$ was an elegant way to avoid this problem \cite{Glashow:1970gm}.
The charmed quark also was crucial in avoiding ``anomalies,'' effects due
to triangle diagrams involving internal fermions and three external gauge
bosons \cite{Bouchiat:1972iq,Georgi:1972bb,Gross:1972pv}.
Evidence for charm was first found in 1974 in the form of the $J/\psi$
particle \cite{Aubert:1974js,Augustin:1974xw}, a bound state of $c$ and $\bar
c$.  An earlier Resource Letter \cite{Rosner:1980hw} deals with events leading
up to this discovery, as well as early evidence for the fifth ($b$) quark
to be mentioned below.  The whole topic of electroweak unification is dealt
with at an intermediate level in several references mentioned earlier
(e.g., \cite{Okun:1982ap,Quigg:1997,Gottfried:1984ck}).

\leftline{\bf F.  CP violation}
\bigskip

The symmetries of time reversal (T), charge conjugation (C), and space
inversion or parity (P) have provided both clues and puzzles in our
understanding of the fundamental interactions.  The realization that the
charge-changing weak interactions violated P and C maximally was central to
the formulation of the $V-A$ theory.  The theory was constructed in 1957 to
conserve the product CP, but the discovery in 1964 of the long-lived neutral
kaon's decay to two pions ($K_L \to \pi \pi$) \cite{Christenson:1964fg} showed
that even CP was not
conserved.  In 1973, Kobayashi and Maskawa (KM) \cite{Kobayashi:1973fv}
proposed that CP violation in the neutral kaon system could be explained in a
model with three families of quarks.  The quarks of the third family, now
denoted by $b$ for bottom and $t$ for top, were subsequently discovered in 1977
\cite{Herb:1977ek,Innes:1977ae} and 1994 \cite{Abe:1994st,Abe:1994xt,%
Abe:1995hr,Abachi:1994je,Abachi:1995td,Abachi:1995iq}, respectively.
Popular articles on these discoveries include one by Lederman
\cite{Lederman:1978gi} and Liss and Tipton \cite{Liss:1997vk}.

An alternative theory of CP violation in the kaon system, proposed by
Wolfenstein \cite{Wolfenstein:1964ks}, involved a ``superweak'' CP-violating
interaction mixing $K^0$ and $\bar K^0$, which would lead to identical CP
violation in $K_L \to \pi^+ \pi^-$ and $K_L \to \pi^0 \pi^0$.  The discovery
that this was not so (see \cite{Alavi-Harati:1999xp,Lai:2001ki} for the most
recent published results, which are continually being updated in conference
reports) disproved the superweak theory and displayed a ``direct'' form
of CP violation with magnitude consistent with that predicted by the
KM theory.

Decays of hadrons containing $b$ quarks are further ground for testing the KM
hypothesis and for displaying evidence for new physics beyond this ``standard
model'' of CP violation.  A meson containing a $\bar b$ quark will be known
generically as a $B$ meson.  Electron-positron colliders have been constructed
at SLAC (Stanford, CA) \cite{Aubert:2000bz} and KEK (Tsukuba, Japan)
\cite{Bondar:2001ud} expressly to study $B$ mesons; others at DESY (Hamburg,
Germany) and Cornell (Ithaca, NY) \cite{Shibata:1988uh} were fortunate in
having just the right energy to produce $B$ mesons in pairs.  The BaBar
detector at SLAC and the Belle detector at KEK have already produced a series of
major results on B decays and CP violation \cite{Aubert:2002ic,Abe:2002bx}.
Studies of particles containing $b$ quarks also are expected to be an important
part of the physics program at the Fermilab Tevatron \cite{Anikeev:2001rk} and
the CERN Large Hadron Collider (LHC) \cite{Schopper:2001ur}.

\leftline{\bf G.  Dynamics of heavy quarks}
\bigskip

With the discovery of the charmed (Sec.\ IX E) and beauty (Sec.\ IX F)
quarks, a whole new laboratory emerged for the study of QCD.  A bound
state of a heavy quark and its antiquark, $c \bar c$ or $b \bar b$, is
known as {\it quarkonium}, in analogy with positronium, the bound state
of a positron and an electron.  (The top quark lives too short a time for
$t \bar t$ bound states to be of much interest, though one can study some
effects of the binding.)  Quarkonium states have been extensively studied,
\cite{Appelquist:1978aq,Novikov:1978dq,Kwong:1987mj,Buchmuller:1988nq},
with their spectroscopy and decays providing useful information on QCD at
various distance scales.

The states of light quarks bound to a single heavy quark have their own
regularities.  They are analogous to atoms in which the light quarks and
gluons represent the ``electronic'' degrees of freedom, while the heavy
quarks represent the nuclei.  Thus, certain properties of these states are
related in the same way that, for example, properties of hydrogen and
deuterium are related.  This ``heavy quark symmetry'' \cite{Isgur:1992xa} has
provided very useful guides to the properties of hadrons containing charm
and beauty quarks, and permits more precise determinations of underlying
weak couplings (such as elements of the Cabibbo-Koyayashi-Maskawa [CKM]
matrix).
\bigskip

\leftline{\bf H.  Higgs boson(s)}
\bigskip

An unbroken \ew~theory involving the photon would require {\it all} fields to
have zero mass, whereas the $W^\pm$ and $Z$ are massive.  The symmetry-breaking
that generates $W$ and $Z$ masses must not destroy the renormalizability of 
the theory.  The {\it Higgs mechanism} achieves this goal at the price of
introducing an additional degree of freedom correponding to a physical
particle, the {\it Higgs particle}, which is the subject of intense searches
\cite{Gunion:1990,Veltman:1986br,Carena:2000yx,Cavalli:2002vs}.  Current
95\% c.l.\ limits on a standard-model Higgs boson are $M_H > 114$ GeV/$c^2$ via
direct searches \cite{LEPHWG} and $M_H < 193$ GeV/$c^2$ from fits to
precise electroweak data \cite{LEPEWWG}.

Discovering the nature of the Higgs boson is a key to further progress in
understanding what may lie beyond the Standard Model.  There may exist one
Higgs boson or more than one.  There may exist other particles in the spectrum
related to it.  The Higgs boson may be elementary or composite.  If composite,
it points to a new level of substructure of the elementary particles.
\bigskip

\leftline{\bf I.  Precision electroweak measurements}
\bigskip

Precision electroweak measurements can yield information on many
new-physics possibilities in addition to the Higgs boson.  The seminal
paper of Veltman \cite{Veltman:1977kh} showed how the ratio of $W$ and $Z$
masses could shed light on the top quark's mass.  A systematic study of
electroweak radiative corrections within the Standard Model was performed
by Marciano and Sirlin \cite{Marciano:1980pb} and used to analyze a wide
variety of electroweak data, initially in \cite{Amaldi:1987fu} and most
recently in \cite{LEPEWWG}.  Widely-used parametrizations of deviations
from Standard-Model predictions \cite{Peskin:1990zt,Peskin:1992sw,%
Altarelli:1991zd} have been used to constrain new particles in higher-order
loop diagrams associated with $W$, $Z$, and photon self-energies.  Some
reviews include Refs.\ %
\cite{Aoki:1982ed,Hollik:1992qb,Bardin:1999ak,Anlauf:2001}.

\leftline{\bf X.  PROPOSED EXTENSIONS}
\bigskip

\leftline{\bf A.  Supersymmetry}
\bigskip

Unification of the electroweak and strong interactions at a high mass scale
leads to the {\it hierarchy problem}, in which this scale contributes through
loop diagrams to the Higgs boson mass and requires it to be fine-tuned at each
order of perturbation theory.  A similar problem is present whenever there is
a large gap between the electroweak scale and {\it any} higher mass scale
contributing to the Higgs boson mass. {\it Supersymmetry} solves this problem by
introducing for each particle of spin $J$ a {\it superpartner} of spin $J \pm
1/2$ whose contribution to such loop diagrams cancels the original one in the
limit of degenerate masses.  Recent reviews of supersymmetry and its likely
experimental signatures include \cite{Ramond:1999vh,Martin:1997ns,%
Ambrosanio:1998zf,%
Abel:2000vs,Culbertson:2000am,Allanach:2002nj,Kane:2002tr}, while
earlier discussions are given by \cite{Nilles:1984ge}, \cite{Haber:1985rc}, and
\cite{Sohnius:1985qm}.  For an article at the popular level see
\cite{Haber:1986ht}.
\bigskip

\leftline{\bf B.  Dynamical electroweak symmetry breaking}
\bigskip

If the Higgs boson is not fundamental but arises as the result of a new
super-strong force which, in analogy with color, causes the {\it dynamical}
generation of one or more scalar particles, the hierarchy problem can be
avoided.  This scheme, sometimes called ``technicolor,'' was proposed in
the 1970s \cite{Weinberg:1976gm,Weinberg:1979bn,Susskind:1979ms}.
For recent reviews, see. e.g.,
\cite{Chivukula:2000mb,Lane:2002wv,Hill:2002ap}.

\leftline{\bf C.  Fermion mass and mixing patterns}
\bigskip

The transitions between the $(u,c,t)$ and $(d,s,b)$ quarks owing to virtual $W$
emission or absorption are described by the Cabibbo-Kobayashi-Maskawa (CKM)
matrix mentioned in Sec.\ IX A.  (For one parametrization of this matrix see
\cite{Wolfenstein:1983yz}.)  The CKM matrix arises because the matrices that
diagonalize the mass matrices of $(u,c,t)$ and of $(d,s,b)$ are not the same.
A theory of quark masses would thus entail a specific form of the CKM matrix.
For the corresponding matrix for leptons, see \cite{Maki:1962mu,Lee:1977qz,%
Lee:1977ti}.  While a theory of quark and lepton masses
still eludes us, attempts have been made to guess some of its general features
\cite{Fritzsch:1978vd,Froggatt:1979nt,Barbieri:1997ww,Ramond:1998pp,%
Fritzsch:1999ee,Albright:2001xq}.

\leftline{\bf D.  Composite quarks and leptons}
\bigskip

Families of quarks and leptons appear to be replicas of one another (see
Table \ref{tab:ql}), aside from their differing masses and weak couplings.
Attempts have been made to explain this regularity in terms of a composite
structure, much as the periodic table of the elements reflects their underlying
atomic structure.  A set of guidelines for this program was laid down by
't Hooft \cite{'t Hooft:1979bh}.  For an example of a recent effort, see
\cite{Arkani-Hamed:1998fq}.

\leftline{\bf E.  Grand unification and extended gauge groups}
\bigskip

An early point in favor of quark--lepton unification was the anomaly
cancellation \cite{Bouchiat:1972iq,Georgi:1972bb,Gross:1972pv} mentioned in
Sec.\ IX E.  The idea that lepton number could be regarded as a fourth
``color,'' leading to an extended gauge group embracing both electroweak and
strong interactions, was proposed by Pati and Salam \cite{Pati:1973uk}.

The strong and electroweak coupling constants are expected to approach one
another at very small distance (large momentum) scales \cite{Georgi:1974yf},
suggesting {\it grand unified theories} based on symmetry groups such as
SU(5) \cite{Georgi:1974sy}, SO(10) \cite{Georgi:1975my}, and E$_{\rm 6}$
\cite{Gursey:1976ki}.  (For an early popular article on this program see
\cite{Georgi:1981yp}.)  These theories typically predict that the proton
will decay \cite{Weinberg:1981kk,Losecco:1985xj,Langacker:1992qb}, and some of
them entail additional observable gauge bosons besides those of the SU(3)
$\times$ SU(2) $\times$ U(1) Standard Model \cite{Hewett:1989xc}.  Some useful
group-theoretic techniques for model-building are described in
\cite{Slansky:1981yr}.

\leftline{\bf F.  Strong CP problem and axions}
\bigskip

In a non-abelian gauge theory such as SU(3) there can arise non-trivial
gauge configurations that prevent terms in the Lagrangian proportional to
Tr ($G_{\mu \nu} \tilde G^{\mu \nu})$ from being ignored as pure divergences.
Such terms can lead to strong CP violation.  Their coefficient, a parameter
conventionally called $\theta$, must be of order $10^{-10}$ or smaller in order
not to conflict with limits on the electric dipole moment of the neutron
\cite{Harris:1999jx}.  Several proposals have been advanced for why $\theta$ is
so small \cite{Ramond:1999vh,Dine:2000cj}.
In one of the most interesting, $\theta$ is promoted
to the status of a dynamical variable that can relax to a natural value of
zero.  As a consequence, there arises a nearly massless particle known as
the {\it axion}, whose properties (and the search for which) are
well-described in \cite{Ramond:1999vh,Dine:2000cj}.

\leftline{\bf G.  String theory}
\bigskip

A truly unified theory of interactions must include gravity.  The leading
candidate for such a theory is {\it string theory}, which originated in
pre-QCD attempts to explain the strong interactions \cite{Nambu:1970,%
Fubini:1969wp,Susskind:1970xm,Nambu:1974zg} by replacing the space-time
points of quantum field theories with extended objects (``strings'').
In 1974 it was realized that string theories necessarily entailed a massless
spin-2 particle, for which the graviton was an ideal candidate
\cite{Scherk:1974ca}.  While it appeared that such theories required
space-time to be 26-dimensional (or 10-dimensional in the presence of
supersymmetry), these extra dimensions were interpreted in the 1980s as a
source of the internal degrees of freedom characterizing particle quantum
numbers (see. e.g., \cite{Gross:1985dd,Gross:1985fr,Gross:1986rr}).
A typical scenario whereby string theory might yield predictions for the
quark and lepton spectrum is described in \cite{Candelas:1985en}.

Early results on string theory are described in the textbook by Green,
Schwarz, and Witten \cite{Green:1987sp,Green:1987mn}.  Later texts are
\cite{Polchinski:1998rq,Polchinski:1998rr}.  Descriptions for the
non-specialist are given by Green \cite{Green:1986xd}, Duff \cite{Duff:1998ec},
Greene \cite{Greene:1999kj} and Weinberg \cite{Weinberg:1999az}.

\leftline{\bf H.  Large extra dimensions}
\bigskip

Although the usual superstring scenario envisions the six extra dimensions
in such theories as having spatial extent of the order of the Planck scale,
$(G_N \hbar/c^3)^{1/2} \simeq 10^{-33}$ cm, theories have been proposed in
which some of the extra dimensions are larger, leading to observable effects at
accelerators or in precise tests of Newton's universal inverse square law of
gravitation \cite{Antoniadis:1990ew,Lykken:1996fj,Arkani-Hamed:1998rs,
Antoniadis:1998ig,Arkani-Hamed:1998nn}.  Reviews for the non-specialist have
appeared in {\it Scientific American} \cite{Arkani-Hamed:2000} and {\it
Physics Today} \cite{Arkani-Hamed:2002rz}.

\leftline{\bf XI.  HINTS OF NEW PHYSICS}
\bigskip

\leftline{\bf A.  Neutrino masses}
\bigskip

The ability of neutrinos of one species to undergo oscillations into another
is an indication of non-zero and non-degenerate neutrino masses
\cite{Bilenky:1987ty,Kayser:2001ki}. Several experiments find evidence for such
oscillations.  Reviews have appeared in \cite{Jung:2001,Conrad:1998ne,%
Kearns:1999kr}; the second of these also deals with precision electroweak tests
using neutrinos.
\medskip

\leftline{\it 1.  Solar neutrinos:}
Since the earliest attempts to detect neutrinos originating from the Sun in the
mid-1960s, the flux has been less than predicted in the standard solar model
\cite{Bahcall:1990qf}.  Recent experiments at the Sudbury Neutrino Observatory
(SNO) in Ontario \cite{Ahmad:2001an,Ahmad:2002jz} and the KamLAND experiment
in Japan \cite{KamLAND:2002}
strongly suggest that this deficit is due to oscillations of the electron
neutrinos produced in the Sun into other species, most likely a combination of
muon and tau neutrinos, induced by interaction with the Sun in a manner (now
known as the MSW effect) first proposed by Mikheev and Smirnov
\cite{Mikheev:1985gs} and Wolfenstein \cite{Wolfenstein:1978ue}.  For
reviews, see \cite{Bahcall:2002n,Bahcall:2002}.
\medskip

\leftline{\it 2.  Atmospheric neutrinos:}
Neutrinos produced by the interactions of cosmic rays in the atmosphere are
expected to be in the ratio $\nu_\mu:\nu_e = 2:1$ (summing over neutrinos
and antineutrinos) \cite{Gaisser:2002jj}.
Instead, a ratio more like 1:1 is observed.  This
phenomenon has been traced to oscillations that are most likely $\nu_\mu
\to \nu_\tau$, as a result of definitive experiments performed by the
Super-Kamiokande Collaboration in Japan \cite{Fukuda:1998mi,Fukuda:2000np}.
The mixing appears to be close to maximal, in contrast to the small
mixings of quarks described by off-diagonal elements of the CKM matrix.
\medskip

\leftline{\it 3.  Indications in an accelerator experiment:}
An experiment performed at Los Alamos National Laboratory \cite{Aguilar:2001ty}
in the Liquid Scintillator Neutrino Detector (LSND) finds evidence for $\bar
\nu_\mu \to \bar \nu_e$ oscillations.  An experiment known as MiniBooNE
which has begun to operate at Fermilab will check this possibility
\cite{Hawker:2001pt}.

\leftline{\bf B.  Cosmic microwave background radiation}
\bigskip

The 2.7 K radiation left over from the Big Bang contains a wealth of
information about both the early Universe and particle physics.  In particular,
the spatial pattern of its fluctuations indicates that the Universe is exactly
on the border between open and closed, and strongly supports the idea that
the Universe underwent a period of exponential inflation early in its
history \cite{Caldwell:2001kq,Gibbs:2002js,Olive:2002,Fukugita:2002,%
Smoot:2002}.  For a review of the cosmological parameters, see
\cite{Bahcall:1999xn}.

\leftline{\bf C.  Baryon asymmetry of the Universe}
\bigskip

To explain why the visible Universe seems to contain so many more
baryons than antibaryons, Sakharov \cite{Sakharov:1967dj} proposed shortly
after the discovery of CP violation that three ingredients were needed:
(1) CP (and C) violation; (2) baryon number violation, and (3) a period in
which the Universe is not in thermal equilibrium.  All of these conditions
are expected to be satisfied in a wide range of theories, such as grand
unified theories (Sec.\ XI.E) in which quarks and leptons, and the electroweak
and strong interactions, are unified with one another \cite{Kolb:1983ni}.
However, details of the mechanism are not clear \cite{Wilczek:1980vy,%
Quinn:1998ru}.  In some
versions of the theory, for example, it is lepton number that is violated in
the early stages of the Universe, giving rise to a lepton asymmetry that is
then converted to a mixture of lepton and baryon asymmetry when the Universe
has evolved further.  For a recent review of this suggestion, see
\cite{Buchmuller:2000as}.

\leftline{\bf D.  Dark matter}
\bigskip

Only a small fraction of the matter in the Universe can be accounted for by
baryons, leaving the remainder to consist of as-yet-unidentified matter
or energy density \cite{Krauss:1986mj}.  Candidates for this {\it dark matter}
are discussed in the {\it Review of Particle Physics} \cite{Srednicki:2002}.
One class of candidates consists of the lightest supersymmetric particle (LSP),
which may be stable; these suggestions are reviewed in \cite{Jungman:1996df}.

\leftline{\bf E.  Dark energy}
\bigskip

The Universe appears not only to be expanding, but its expansion appears to be
speeding up.  Evidence for this behavior comes from the study of distant
supernovae, which furnish ``standard candles'' for a cosmological distance
scale \cite{Hogan:1999mx,Bahcall:1999xn}.  One interpretation is that a {\it
cosmological
constant} $\Lambda$ (first proposed by Einstein shortly after he formulated
the general theory of relativity) accounts for about 65\% of the energy
density of the Universe.  This contribution is sometimes referred to as ``dark
energy,'' to distinguish it from the ``dark matter'' accounting for nearly all
of the remaining energy density aside from a few-percent contribution from
baryons \cite{Olive:2002,Fukugita:2002}.  An alternative suggestion is that
the ``dark energy'' is due to a new field, dubbed ``quintessence''
\cite{Ostriker:2001ik}.  For recent accounts of ``dark energy'' see
\cite{Kirshner:2002} and \cite{Peebles:2002gy}.

\bigskip

\leftline{\bf XII.  EXPERIMENTAL APPROACHES}
\bigskip

The rise of the Standard Model would not have been possible without a variety
of experimental facilities, including accelerators, detectors, and
non-accelerator experiments.  What follows is a brief description of some
currently operating laboratories and experiments.  Fuller descriptions may be
found through laboratory web sites, listed in Sec.\ VII.B, and through web
sites of specific collaborations.  Some references to recent experiments are
given in this Section.
\bigskip

\leftline{\bf A.  High energy accelerator facilities}
\bigskip

\leftline{\it 1. Beijing Electron-Positron Collider (China)}
This electron-positron collider with center-of-mass energy 2--5 GeV recently
reported an improved measurement of $R$ (see Sec.\ II.C) in this energy range
\cite{Bai:2001ct}.  It has made important contributions to the study of
$\tau$ leptons, charmed particles, and $c \bar c$ bound states.
\medskip

\leftline{\it 2.  Brookhaven National Laboratory (U.S.A.)}
The {\it Alternating-Gradient Synchrotron} (AGS) is a fixed-target proton
accelerator with maximum energy of about 30 GeV.  The first neutrino beam
constructed at an accelerator was used at the AGS to show that the muon and
electron neutrino are distinct from one another \cite{Danby:1962nd}. 
One of its most spectacular
discoveries was the $J/\psi$ particle, a bound state of a charmed quark and a
charmed antiquark \cite{Aubert:1974js}.  Recent experiments include
the detection of the rare process $K^+ \to \pi^+ \nu \bar \nu$
\cite{Adler:2001xv} and a precise measurement of the muon anomalous magnetic
moment \cite{Brown:2001mg}.  It serves as an injector to the {\it
Relativistic Heavy-Ion Collider} (RHIC), whose maximum energy of about 200
GeV per nucleon permits studies of the quark-gluon plasma and other aspects
of hadron physics at high densities.
\medskip

\leftline{\it 3. CERN (Switzerland and France)}
CERN's 28-GeV Proton Synchrotron (PS) began operation in 1959.  It served as a
source of protons for the Intersecting Storage Rings (ISR), which began
operation in the early 1970s and achieved a maximum center-of-mass energy of 62
GeV.  Its protons were used to produce neutrinos which provided the first
evidence for neutral currents in 1973 \cite{Hasert:1973cr,Hasert:1973ff}.
The {\it Super-Proton-Synchrotron} (SPS), a 400-GeV fixed-target
machine built in the mid-1970s, was converted to a proton-antiproton collider
(the ``S$\bar p p$S'') early in the 1980s, leading to the discovery of the $W$
and $Z$ bosons in 1983 \cite{Rubbia:1984,VanDerMeer:1984}.  The Large
Electron-Positron (LEP) Collider \cite{Myers:1990gu} was commisioned in 1989,
making a series of precise measurements at the center-of-mass energy of the $Z$
boson (91.2 GeV) (an early measurement of the $Z$ width pointed to three
families of quarks and leptons \cite{Feldman:1991kw}) before moving up in
energy to nearly 210 GeV and ending its program in 2000 \cite{Drees:2001xw}.
Its magnets have been removed, making way for the Large Hadron Collider (LHC),
a proton-proton collider that will have a c.m.\ energy of 14 TeV
\cite{LlewellynSmith:2000nb,Foa:1999vm}.
\bigskip

\leftline{\it 4. CLEO/CESR at Cornell (U.S.A.)}
The Wilson Synchrotron at Cornell, a circular electron accelerator built in
1967, was converted in 1979 to an electron-positron collider, the Cornell
Electron Storage Ring (CESR), with maximum energy 8 GeV per beam
\cite{Berkelman}.  It arrived
on the scene just in time to study the $\Upsilon(1S)$ $b \bar b$ resonance and
its excited states, including the $\Upsilon(4S)$ which decays to a $B \bar B$
meson pair.  Studies of $B$ mesons have dominated the program of the CLEO
detector at CESR until recently.  For the next year or two, CLEO will return
to the $\Upsilon(1S,2S,3S)$ resonances, after which it is planned to
optimize CESR to run at the lower energies appropriate for charm production
\cite{Shipsey:2002kc}.  This will permit a return to many interesting questions
with a vastly improved detector and statistical sample.
\bigskip

\leftline{\it 5. DESY (Germany)}
A circular electron accelerator at the Deutsches Elektronen Synchrotron (DESY)
laboratory was converted to an electron-positron collider (DORIS) whose
experimental program paralleled that of CESR/CLEO for a number of years,
yielding important information about $\Upsilon$ spectroscopy and $B$ mesons,
for example through work of the ARGUS Collaboration.  Subsequent machines
included the larger $e^+ e^-$ collider PETRA (maximum c.m.\ energy 46 GeV)
and the currently operating HERA lepton-proton collider, which has studied
both $e^-p$ and $e^+ p$ interactions.  HERA has extended information on
deep inelastic lepton scattering to new kinematic regimes and provided
important information on the gluon structure of the proton.
\bigskip

\leftline{\it 6. Fermilab (U.S.A.)}
The Fermi National Accelerator Laboratory in Batavia, Illinois, U.S.A., began
operation in 1972 as a proton accelerator with initial energy 200 GeV,
rising to 400 GeV within a year.  With the addition of a ring of
superconducting magnets in 1983 it was converted to an energy of 800 GeV
capable of providing protons to fixed targets and proton-antiproton
collisions with a center-of-mass energy of 1.8 TeV \cite{Lederman:1991tm,%
Hoddeson:1987}.  Its energy has recently
been upgraded to nearly 1 TeV per beam with the addition of a new 150-GeV
proton ring called the Main Injector.  Outstanding discoveries at Fermilab
include those of the bottom quark in 1977 \cite{Herb:1977ek,Innes:1977ae},
the top quark in 1994 \cite{Abe:1994st,Abe:1994xt,Abe:1995hr,Abachi:1994je,%
Abachi:1995td,Abachi:1995iq}, and the tau neutrino in 2000
\cite{Kodama:2000mp}.
\bigskip

\leftline{\it 7. Frascati (Italy)}
A major pioneer in the study of electron-positron collisions has been the
Laboratori Nazionali di Frascati (INFN) near Rome, Italy.  Starting in the
early 1960s with the ADA collider and continuing through the ADONE storage
ring, which begain operation in the late 1960s, the laboratory has now
begun to operate a machine called DA$\Phi$NE (DAFNE), which seeks to
produce kaons and other particles through the reaction $e^+ e^- \to \phi
\to \ldots$ at a center-of-mass energy of 1.02 GeV.
\bigskip

\leftline{\it 8. KEK (Japan)}
In the early 1970s, a 12-GeV proton synchrotron was constructed in Japan
near Tokyo at the National Laboratory for High Energy Physics, for which KEK
(Ko-Energi-Kenkyujo) is the acronym in Japanese.  The next major project
at KEK, the TRISTAN $e^+e^-$ collider, attained a center-of-mass energy
in excess of 60 GeV, the highest in the world for such a machine at its debut
in 1986.  Among the topics studied by TRISTAN included weak--electromagnetic
interference through the processes $e^+ e^- \to (\gamma^*,Z^*) \to \ldots$,
where the asterisk denotes a virtual photon or $Z$.  The latest project at
KEK is the KEK-B $e^+e^-$ collider, a lower-energy machine built in the
TRISTAN tunnel, which is designed to produce pairs of $B$ mesons with net
motion on their center-of-mass by using unequal electron and positron energies.
In this way the positions at which the $B$ mesons decay can be spread out
longitudinally, permitting easier study of time-dependences that are of
particular interest in CP-violating processes.  The Belle detector operating
at KEK-B \cite{Bondar:2001ud} is producing significant results on $B$ decays,
as mentioned above \cite{Abe:2002bx}), as is the BaBar
detector operating at PEP-II (see the description of SLAC, below).
\bigskip

\leftline{\it 9. Novosibirsk (Russia)}
A series of $e^+ e^-$ colliders has operated at the Budker
Institute for High Energy
Physics in Novosibirsk for a number of years.  Indeed, work at this laboratory
helped to pioneer the study of beam dynamics essential for achieving such
collisions.  These colliders performed important measurements at the
center-of-mass energies of the $\Upsilon(9.46)$ and $\phi(1.02)$ resonances,
where the numbers denote the mass in GeV/$c^2$.
\bigskip

\leftline{\it 10.  Protvino (Russia)}
The largest accelerator at present in Russia is a 76-GeV proton synchrotron
at Serphukhov (Protvino), which began operation in the early 1970s.  It was
the first to detect rising meson-baryon cross sections \cite{Denisov:1971jb},
followed soon by the observation of a similar effect in proton-proton
collisions at the CERN ISR (see above).
\bigskip

\leftline{\it 11. SLAC (U.S.A.)}
The early program of the 30-GeV 2-mile-long linear electron accelerator at the
Stanford Linear Accelerator Center (SLAC) included the discovery of pointlike
constituents inside the proton through deep inelastic scattering
\cite{Riordan:1987gw,Bloom:1969kc,Breidenbach:1969kd}.  In the early 1970s the
SPEAR electron-positron storage ring was constructed with maximum
center-of-mass energy equal to 7.4 GeV.  Late in 1973 this machine confirmed a
surprising enhancement of the $e^+ e^-$ annihilation cross section starting at
a c.m.\ energy of 4 GeV seen earlier at the Cambridge Electron Accelerator
(CEA), and in 1974 was one of two sources of the discovery of the $J/\psi$
particle \cite{Augustin:1974xw}, the other being a fixed-target experiment at
Brookhaven National Laboratory \cite{Aubert:1974js} (see above).  In the
mid-1970s construction was begun on PEP, an
electron-positron collider with c.m.\ energy of about 30 GeV, which performed
studies of the elctroweak theory and was the first to measure the $b$ quark
lifetime.  The energy of the LINAC was then raised to 50 GeV, both electrons
and positrons were accelerated, and these were then bent in arcs to collide
with one another at energies equal to or greater than the mass of the $Z$
boson.  This machine, the Stanford Linear Collider (SLC) \cite{Rees:1989zb},
pioneered in precision studies of the $Z$ boson through its Mark II and SLD
detectors; its early measurement of the $Z$ width was a piece of evidence
for three families of quarks and leptons \cite{Feldman:1991kw}.  The latest
SLAC project, the PEP-II asymmetric $e^+ e^-$ collider, has seen evidence for
CP violation in $B$ decays in its BaBar detector \cite{Aubert:2000bz,Aubert:%
2002ic} (see also KEK-B and Belle, above), and has achieved record luminosity
for any collider.  By the middle of this decade both BaBar and Belle expect to
have produced and recorded several hundred million $B \bar B$ pairs.
\bigskip

\leftline{\it 12.  Thomas Jefferson National Laboratory (U.S.A.)}
A moderate-energy (5.7-GeV) electron accelerator, this machine studies
interactions with nuclei and the photoproduction and
electroproduction of resonances containing light quarks ($u,d,s$), with an
eye to seeing those that cannot be explained purely as $q \bar q$ mesons
or $qqq$ baryons.  An upgrade to 12 GeV is under discussion.

\leftline{\bf B.  Non-accelerator experiments}
\bigskip

\leftline{\bf 1.  Underground or underwater laboratories}
\bigskip

The ability to perform experiments in a low-background environment is
greatly increased by going deep underground, where cosmic ray interactions
are less frequent.  A number of major laboratories now are operating
underground, including ones at the Kamioka mine (Japan) \cite{Kamioka}, Gran
Sasso (Italy) \cite{GS}, and Soudan (Minnesota, U.S.A.) \cite{Soudan}.  Whereas
the focus of several laboratories initially had been the search for proton
decay, it has now broadened to include the study of interactions of neutrinos
from atmospheric cosmic rays, the Sun, and even supernovae, and the search
for effects of dark matter.

The next stage of operation of detectors in the laboratories mentioned above
includes the study of artificially produced neutrinos.  The Fermilab
accelerator will send neutrinos to the MINOS detector \cite{MINOS} in Soudan.
The proton synchrotron at KEK in the K2K experiment \cite{Kamioka}, and later a
machine known as the Japan Hadron Facility \cite{JHF}, will direct neutrinos to
the SuperKamiokande detector in Kamioka.  Finally, a detector known as KamLAND
\cite{KamLANDweb}, also in the Kamioka mine, will be sensitive to neutrinos
from reactors over a large portion of Japan, and has already reported its first
results \cite{KamLAND:2002}.

Some current and forthcoming detectors will also be sensitive to naturally
occurring neutrinos.  These include the Sudbury Neutrino Observatory in
Ontario \cite{Sudbury}, the Borexino experiment \cite{Borexino} in Gran Sasso,
and the SuperKamiokande detector mentioned above.  At the South Pole a number
of phototubes have been sunk deep into the ice in the AMANDA experiment
\cite{AMANDA}, which is envisioned in the IceCube experiment \cite{IceCube}
to expand to an effective volume of a cubic kilometer.  The RICE experiment
\cite{RICE} seeks to study the low-frequency tail (at
several hundred MHz) of \v{C}erenkov emission by electrons produced by
neutrinos, also in South Polar ice.  A number of neutrino detectors are also
deployed or planned deep underwater, e.g., in Lake Baikal \cite{Baikal} and the
Mediterranean Sea
(ANTARES \cite{ANTARES}, NEMO \cite{NEMO}, NESTOR \cite{NESTOR}).
\bigskip

\leftline{\bf 2.  Atomic physics}
\bigskip

A large accelerator is not always needed to study fundamental particle
physics beyond the Standard Model.  An example is the window on non-standard
physics provided by atomic parity violation.  (See the bibliography in
\cite{Wieman:2000}.)  Studies of weak-electromagnetic
interfence in atoms such as Cs, Tl, and Pb are in principle sensitive to new
interactions and extended gauge theories, particlarly if the effects of
atomic physics can be separated from more fundamental effects.
\bigskip

\leftline{\bf 3.  Electric and magnetic dipole moments}
\bigskip

The electric dipole moment of the neutron is an excellent probe of physics
beyond the Standard Model, which predicts it to be orders of magnitude
smaller than its current upper bound \cite{Harris:1999jx} of $|d_n| < 6 \times
10^{-26}e \cdot$ cm.  For a bibliography of experimental literature
on electric dipole moments and atomic parity violation, see \cite{Wieman:2000}.

The magnetic dipole moments of particles also provide important constraints
on the Standard Model.  The anomalous magnetic moment of the muon, in 
particular, is sensitive to new-physics effects such as those that arise in
some versions of supersymmetry \cite{Czarnecki:2001pv}.  The current status
of measurements of this quantity indicates a possible deviation from
standard-model predictions, but at a level which is not yet statistically
compelling \cite{Brown:2001mg}.

\leftline{\bf C.  Plans for future facilities}
\bigskip

The particle physics community is developing a number of options to
probe further beyond the Standard Model.  These include a large linear
$e^+ e^-$ collider, intense sources of neutrinos (``neutrino factories''),
a muon collider, and a Very Large Hadron Collider (VLHC) with
energy significantly greater than the LHC.  Descriptions of all of these
options may be found in the Proceedings of the 2001 Snowmass Workshop
\cite{Snowmass2001}.

\leftline{\bf XIII.  SUMMARY}
\bigskip

The Standard Model of electroweak and strong interactions has been in place for
nearly thirty years, but precise tests have entered a phase that permits
glimpses of physics beyond this impressive structure, most likely associated
with the yet-to-be discovered Higgs boson and certainly associated with new
scales for neutrino masses.  Studies of CP violation in decays
of neutral kaons or $B$ mesons are attaining impressive accuracy as well, and
could yield cracks in the Standard Model at any time.  It is time to ask what
lies behind the pattern of fermion masses and mixings.  This is an {\it input}
to the Standard Model, characterized by many free parameters all of which await
explanation.

Many avenues exist for exploration beyond the Standard Model, both theoretical
and experimental.  A lively dialogue between the two approaches must be
maintained, with adequate support for each, if we are to take the next step
in this exciting adventure.
\bigskip

\leftline{\bf ACKNOWLEDGMENTS}
\bigskip

I wish to thank T. Andr\'e, T. Appelquist, R. Cahn, Z. Luo, C. Quigg, G.
Passarino,
R. Shrock, R. Stuewer, O. L. Weaver, and B. Winstein for constructive
comments on the manuscript, and the Theory Group at Fermilab for hospitality.
This work was supported in part by the United States Department of Energy
through Grant No.\ DE FG02 90ER40560.

\end{document}